\documentclass[twocolumn,trackchanges]{aastex63}
\usepackage{amsmath}
\usepackage{graphicx}
\usepackage{amsfonts}
\usepackage{natbib}
\usepackage{color}
\usepackage{epstopdf}
\usepackage{graphicx}
\usepackage{multirow}
\usepackage{longtable}
\usepackage{rotating}

\shorttitle{Transient QPO in the Blazar S4 0954+658}
\shortauthors{Gong et al.}

\newcommand\gr{{$\gamma$-ray}}

\begin{document}

\title{Two Transient Quasi-periodic Oscillations in $\gamma$-Ray Emission from the Blazar S4 0954+658}

\author{Yunlu Gong}
\affil{Department of Astronomy, School of Physics and Astronomy, Key Laboratory of Astroparticle Physics of Yunnan Province, Yunnan University, Kunming 650091, People's Republic of China; fangjun@ynu.edu.cn, gyunlu2021@163.com}

\author{Shiting Tian}
\affil{Department of Astronomy, School of Physics and Astronomy, Key Laboratory of Astroparticle Physics of Yunnan Province, Yunnan University, Kunming 650091, People's Republic of China; fangjun@ynu.edu.cn, gyunlu2021@163.com}

\author{Liancheng Zhou}
\affil{Department of Astronomy, School of Physics and Astronomy, Key Laboratory of Astroparticle Physics of Yunnan Province, Yunnan University, Kunming 650091, People's Republic of China; fangjun@ynu.edu.cn, gyunlu2021@163.com}

\author{Tingfeng Yi}
\affil{Physics Department, Yunnan Normal University, Kunming 650092, People's Republic of China; ytf@ynnu.edu.cn}

\author{Jun Fang}
\affil{Department of Astronomy, School of Physics and Astronomy, Key Laboratory of Astroparticle Physics of Yunnan Province, Yunnan University, Kunming 650091, People's Republic of China; fangjun@ynu.edu.cn, gyunlu2021@163.com}

\begin{abstract}
In this work, we report periodicity search analyses in the gamma-ray light curve of the blazar S4 0954+658 monitoring undertaken by the Fermi Large Area Telescope (LAT). Four analytical methods and a tool are adopted to detect any periodic flux modulation and corresponding significance level, revealing that (i) a 66 d quasi-periodic oscillation (QPO) with the significance level of $> 5\sigma$ spanning over 600 d from 2015 to 2016 (MJD 57145--57745), resulting in continuous observation of nine cycles, which is one of the highest cycles discerned in blazar gamma-ray light curve; (ii) a possible QPO of 210 d at a moderate significance of $\sim3.5\sigma$ lasted for over 880 d from 2020 to 2022 (MJD 59035--59915), which lasted for four cycles. In addition, we discuss several physical models to explain the origin of the two transient QPOs and conclude that a geometrical scenario involving a plasma blob moving helically inside the jet can explain the time scale of the QPO.

\end{abstract}


\section{Introduction}
\label{sec:intro}
It is generally believed that all the active galaxies are powered by the accretion process of dense ionized gases on to the supermassive black hole (SMBH) with a mass in the range of $10^6-10^{10} M_{\odot}$, and $\sim10\%$ of them have relativistic charged particle jets. Radio-loud Active Galactic Nuclei (AGN), with their jets pointing almost directly to observer's line of sight ($\leq 10^{\circ}$), form a special subclass called blazars \citep{1993ARA&A..31..473A,1995PASP..107..803U}. Moreover, blazars can be further divided into two subcategories based on the strength of emission lines emerging in optical-ultraviolet spectra: BL Lacertae objects (BL Lacs; very weak and narrow emission lines) and flat-spectrum radio quasars (FSRQs; broad and strong emission lines). Blazars usually manifest the most substantial variability over almost the whole electromagnetic spectrum and its emission range dominated by nonthermal radiation is from radio  to $\gamma$-rays \citep{1997ARA&A..35..445U}.

Both ground-based and space telescope observations show that the blazars has flux variability of the order of minutes to years at different electromagnetic wavebands, which may indicate that different physical mechanisms (intrinsic and extrinsic) play a leading role. An interesting phenomenon related to flux variability is the QPO, although flux variability frequently exhibit non-linear, stochastic, and aperiodic characteristics \citep{2017ApJ...849..138K}. So far, a large number of QPO behaviors with different timescale in multifrequency light curves have been reported by researchers using different detection techniques \citep[e.g.,][and references therein]{2001A&A...377..396R,2006ApJ...650..749L,2009ApJ...690..216G,2009A&A...506L..17L,King2013,2014RAA....14..933Z,2015Natur.518...74G,
2015ApJ...813L..41A,Bhatta2017,2018A&A...616L...6G,Zhou2018,2020A&A...642A.129S,2022ApJ...925..207Z,2022MNRAS.510.3641R,2022ApJ...931..168G,2023MNRAS.518.5788O}. The detection of QPO phenomenon are usually quite rare and non-persistent for AGNs, but they seem to be relatively common in the black hole X-ray binaries \citep{2006ARA&A..44...49R,2014JApA...35..307G}. So far, more than 30 of 5064 sources above $4\sigma$ significance are reported to have QPO phenomena based on time series data in the fourth Fermi Gamma-ray LAT catalog of sources \citep[4FGL;][]{2020ApJS..247...33A,2022ApJ...929..130W}.

Recently, \cite{2022Natur.609..265J} claimed that the $\gamma$-ray flux, optical flux and linear polarization of BL Lacertae all exhibit $\sim$13 hour QPO variability during a dramatic outburst in 2020. Such a short-term QPO is explained by the current-driven kink instabilities near a recollimation shock $\sim$5 parsecs (pc) from the black hole. In the same year, a quasi periodic signal of approximately 420 days with $>5\sigma$ significance was found in the measurements of the optical linear polarization degree for the blazar PKS 1222+216 and a helical jet model was employed to explain the signal well \citep{2022ApJ...934....3Z}. Furthermore, several models have been proposed by different authors to explain periodic radiation of blazars in various frequencies on diverse time-scales, i.e., a hotspot orbiting near the innermost stable circular orbit of the SMBH \citep{2009ApJ...690..216G,2019MNRAS.484.5785G,2021MNRAS.501...50S}, the presence of a binary system of SMBH \citep{Valtonen2008,2015ApJ...813L..41A}, precession of relativistic jets or helical structure \citep{2015MNRAS.453.1562G,2016AJ....151...54S}, the existence of quasi-equidistant magnetic islands inside the jet \citep{2013RAA....13..705H,2018ApJ...854L..26S,2022MNRAS.510.3641R}, and the pulsational accretion flow instabilities \citep{2018ApJ...854...11T}. Hence, we can analyze the quasi-periodic modulation in the blazar light curve to explore the accretion physics and the connection between accretion disc, jet, and central engine \citep{2020MNRAS.499..653K}.

S4 0954+658 (also referred to as QSO B0954+65) is one of the most well studied source with complex variability in blazars and is situated at a redshift of $z = 0.3694 \pm 0.0011$ \citep{2021MNRAS.504.5258B}. \cite{1991ApJ...374..431S} regard this source as a BL Lac object in view of the small equivalent width of the emission lines of the spectrum. However, this target can also be classified as FSRQs due to the kinematic features of the radio jet belongs to class II \citep{2016A&A...592A..22H}. In 2021, \cite{2021MNRAS.504.5258B} detected a MgII emission line, whose equivalent width is close to 5 Angstrom, commonly taken as the limit to classify blazars as FSRQ. Therefore, it seems more reasonable to consider this \gr\ emitter as a transitional object. \cite{1993A&A...271..344W} investigated the optical variability of this source for the first time and then \cite{1999A&A...352...19R} detected fast large amplitude variations using the 4-yr light curve. Their results indicate that the long term behavior of the source are not related to spectral variations. Then, the continuous observation of this blazar shows that the optical flux variations by more than 2.5 magnitude and a degree of polarization that reached 40\% \citep{2004A&A...426..437P,2015ARep...59..551H}.  Additionally, \cite{2019MNRAS.484.5633G} found a positive correlation between colour index with respect to the magnitude based on the simultaneous data in B and R bands.

In very high-energy ($\ge100$ GeV) $\gamma$-rays, \cite{2018A&A...617A..30M} presented the first detection of the blazar S4 0954+658, which was obtained through monitoring with the Major Atmospheric Gamma Imaging Cherenkov (MAGIC) Telescopes during an exceptional flare (February 2015). In 2021, \cite{2021MNRAS.504.5629R} found a 31.2 day QPO behavior in the optical long-term variability through the observation of the Transiting Exoplanet Survey Satellite (TESS) and the Whole Earth Blazar Telescope (WEBT) Collaboration, in which the rotation of an inhomogeneous helical jet provides a reasonable explanation for this phenomenon. It is worth mentioning that such a month-like transient QPO is also detected in the \gr\ band for PKS 2247-131 \citep{Zhou2018}. More recently, \cite{2023ApJ...943...53K} report the discover of several QPOs around 0.6-2.5 days in the optical light curve of the blazar S4 0954+658 with data acquired in six sectors by the TESS.

Here, we are inspired by the QPO report on the optical radiation, and then try to analyze whether the $\sim$14.3 yr data measured by Fermi-LAT also have QPO phenomenon. The paper is structured as follows. In Section~\ref{data}, we describe the data analysis process of 0.1-300 GeV energy band. In Section~\ref{results}, we present the QPO detection algorithm and main results. In Section~\ref{sumdis}, we summarize our conclusions and explore several models to explain the QPO results.

\begin{figure*}
\centering
\begin{minipage}[t]{0.99\textwidth}
\centering
\includegraphics[height=9cm,width=17cm]{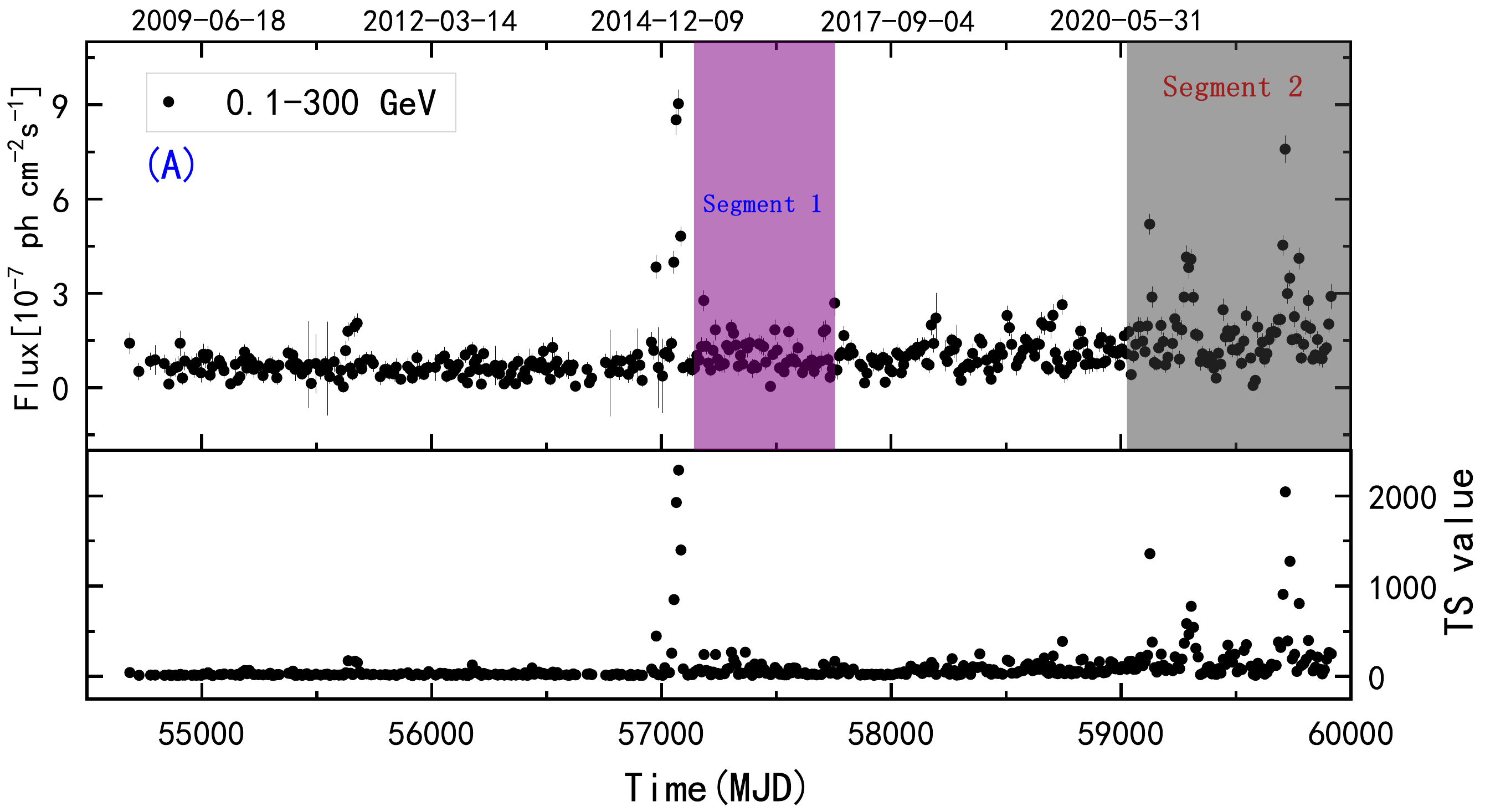}
\end{minipage}
\begin{minipage}[t]{0.99\textwidth}
\centering
\includegraphics[height=6cm,width=17cm]{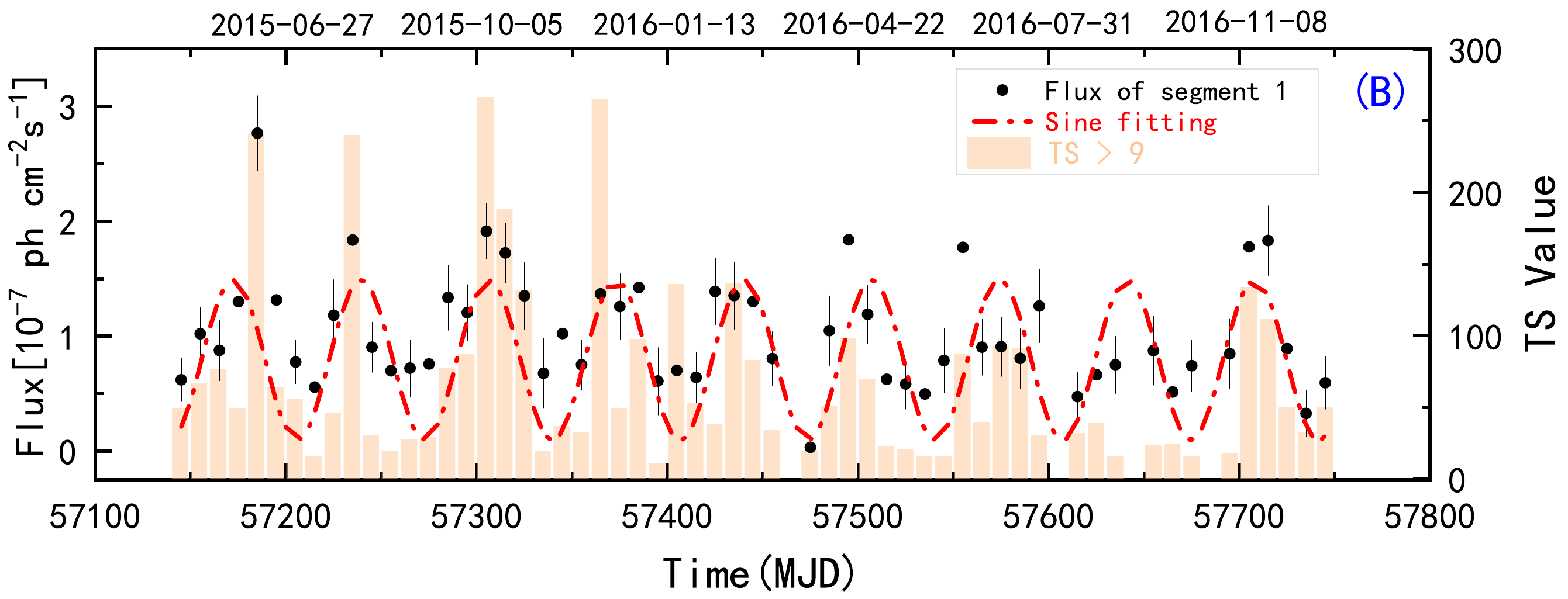}
\end{minipage}
\begin{minipage}[t]{0.99\textwidth}
\centering
\includegraphics[height=6cm,width=17cm]{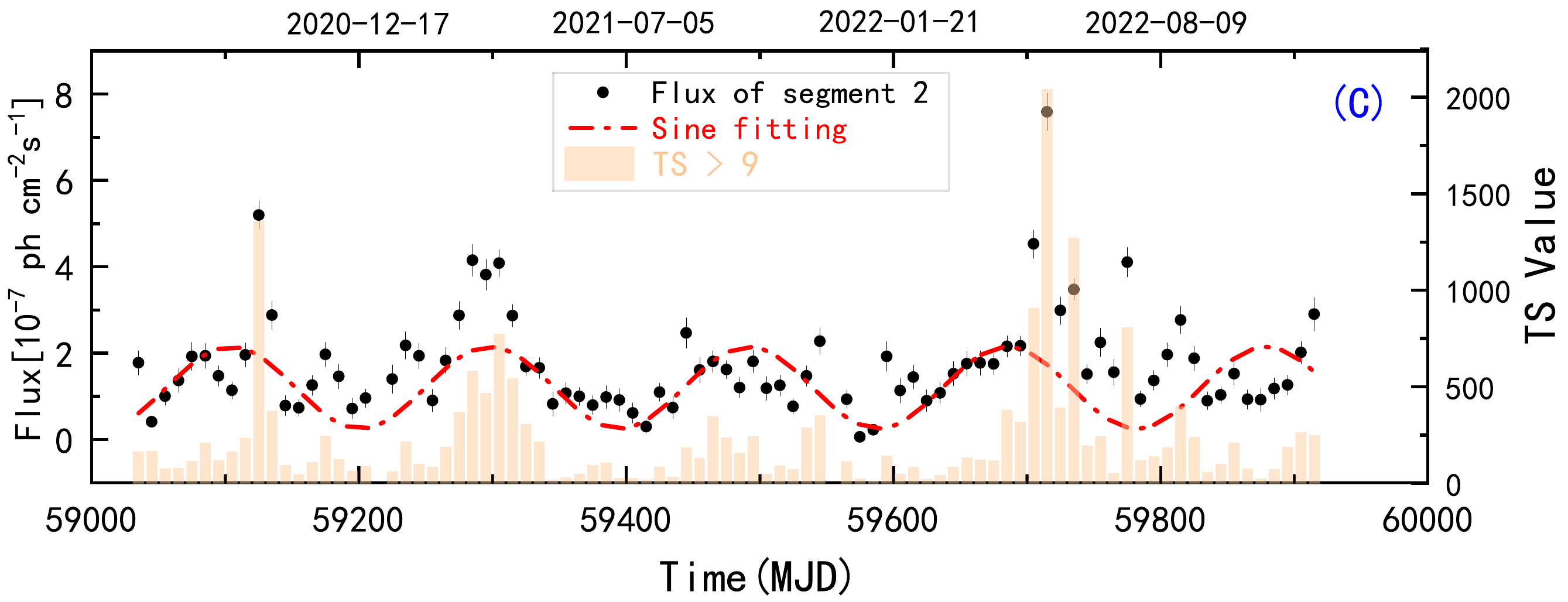}
\end{minipage}
\caption{The 10-day binned light curve of the blazar S4 0954+658 at \gr\ energies of 0.1-300 GeV obtained from Fermi-LAT. Panel A: the purple shaded regions (marked as segment 1) represents the first QPO analysis piece concerned during MJD 57145--57745. The gray shaded region is the epoch MJD 59035--59915 (marked as segment 2) where the QPO analyses were carried out. Panel B: zoom-in of segment 1, where the red dashed-dotted line indicates the sine fitting result of the light curve. The orange histogram corresponds to the TS value of each data point. Panel C: same as panel B but for segment 2.}
\label{Fig1}
\end{figure*}

\section{Fermi-LAT Data Analysis}
\label{data}
The LAT on-board the Fermi observatory continually survey the entire sky every 90 minutes in the energy range from 20 MeV to $>$ 300 GeV \citep{Atwood2009}. Based on the observation data of the first 12 years, the 4FGL incremental version of \gr\ source contains 6658 sources, including more than 100 newly classified blazars \citep{2022ApJS..260...53A}. The blazar S4 0954+658 (named as 4FGL J0958.7+6534) was found in the first Fermi Gamma-ray LAT catalog, and also has been detected by various radio surveys and optical and millimeter surveys. In order to build the light curve of this source, we used the standard software package \texttt{FERMITOOLS} and the user contributed tool \texttt{make4FGLxml.py}.

The data for the blazar S4 0954+658 were taken during the period 2008 August 4 (MET:239557417) to 2022 December 5 (MET:691900553) covering $\sim$14.3 years. We chose LAT 0.1-300 GeV Pass 8 (evclass = 128, evtype = 3) events recommended by the Fermi-LAT collaboration from a circular region of interest having a radius of $12^{\circ}$ centred at the source ($\alpha_{2000.0} = 09^h58^m47.244^s, \delta_{2000.0} = 65^{\circ}33^{\prime}54.8^{\prime\prime}$). At the same time, we used a screening expression $\texttt{``(~DATA\_QUAL > 0)\&\&(~LAT\_CONFIG=1)''}$ to select events with good time intervals and set a zenith angle cut of 90 degrees to suppress the \gr\ pollution from the Earth's limb. An XML file is generated through the 4FGL catalogue containing the \gr\ background emission templates \texttt{`gll\_iem\_v07'} and \texttt{`iso\_P8R3\_SOURCE\_V2\_v1.txt'} for the Galactic and isotropic extragalactic contributions respectively. We consider three commonly used spectral models (power-law, log-parabola, and power law with an exponential cutoff) for the whole time series. And we also test for the spectral curvature in the spectrum using $\Delta TS = TS_{LogPb}-TS_{PL} = 343$ \citep{2020ApJS..247...33A}. The results show that the log-parabola (LogPb) model is more suitable for describing $\gamma$-ray emission of target source. The best-fit spectral parameters were $\alpha = 2.14 \pm 0.01$, $\beta = 0.53 \pm 0.04$, and $E_b = 699.12 \pm 29.46$ MeV. In addition, we selected low (MJD 54687--55558 and 55778--56702) and high states (MJD 56918--57160 and 59619--59894) to test the spectral shape of the time series. The results show that the fitting parameters (except $\beta$) and flux variability are close to the whole time series.

\begin{figure*}
\centering
\begin{minipage}[t]{0.49\textwidth}
\centering
\includegraphics[height=7.0cm,width=8.6cm]{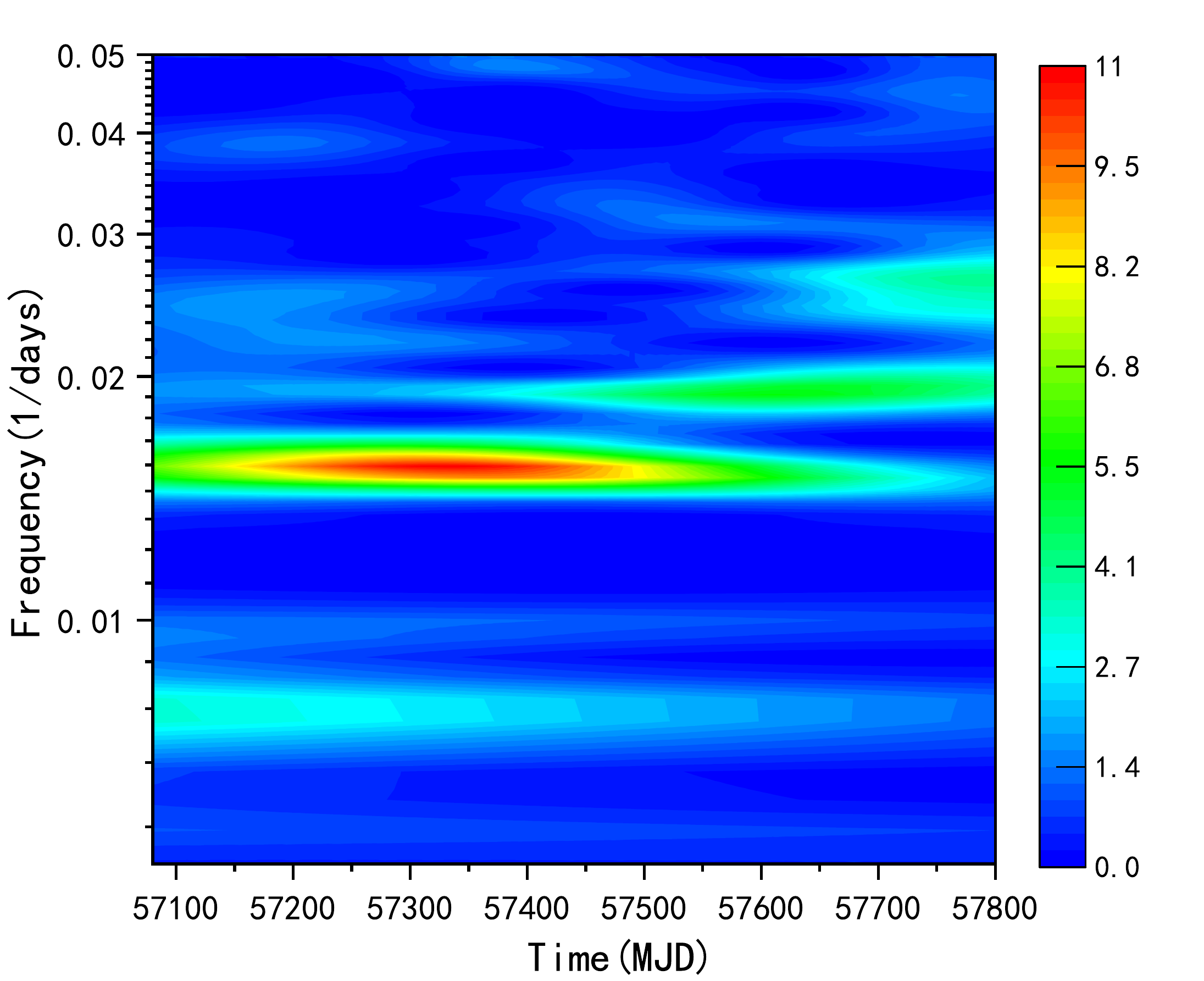}
\end{minipage}
\begin{minipage}[t]{0.49\textwidth}
\centering
\includegraphics[height=7.0cm,width=8.6cm]{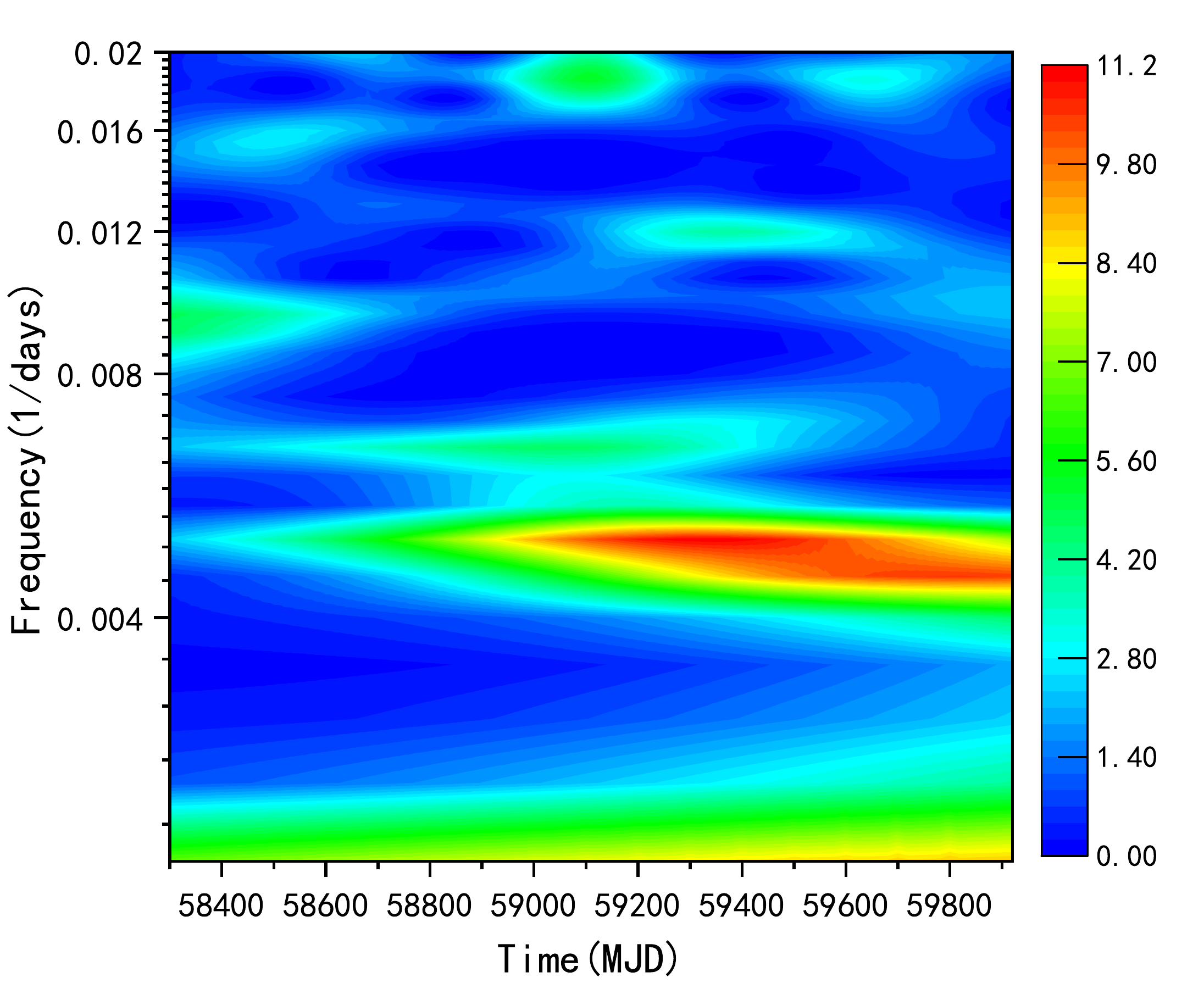}
\end{minipage}
\caption{Left panel: WWZ map of the S4 0954+658 light curve in MJD 57085-57795. The bright red patch represents a possible QPO in the interval of MJD 57145-57745 (segment 1). Right panel: WWZ map of the blazar S4 0954+658 $\gamma$-ray light curve in MJD 58315-59915. The bright red patch represents a possible QPO in the interval of MJD 59035-59915 (segment 2).}
\label{Fig4}
\end{figure*}

Based on the best fitting results mentioned above, we tested the construction of bins light curve for 1-30 days and found that 10 day bins is the most appropriate bin, as they not only reveal the details of the flux variation, but also ensure that the blazar S4 0954+658 can be detected in almost all bins (TS $\geqslant$ 9). In addition, the 10-day binned light curve also shows the strongest intensity in the power spectrum calculation compared with other bins. In the 10-day binned light curve (See Figure~\ref{Fig1}), the average value and standard deviation are 1.11 and 0.95 $\rm\times10^{-7}~photons\,cm^{-2}\,s^{-1}$, respectively. Detection of periodicity in the light curve based on Weighted Wavelet Z-transform (WWZ) method (See Figure~\ref{Fig4}), we selected the panels B (segment 1) and C (segment 2) in Figure~\ref{Fig1} as the regions of interest for QPO variability analysis.

\begin{figure*}
\centering
\begin{minipage}[t]{0.49\textwidth}
\centering
\includegraphics[height=6.5cm,width=7.8cm]{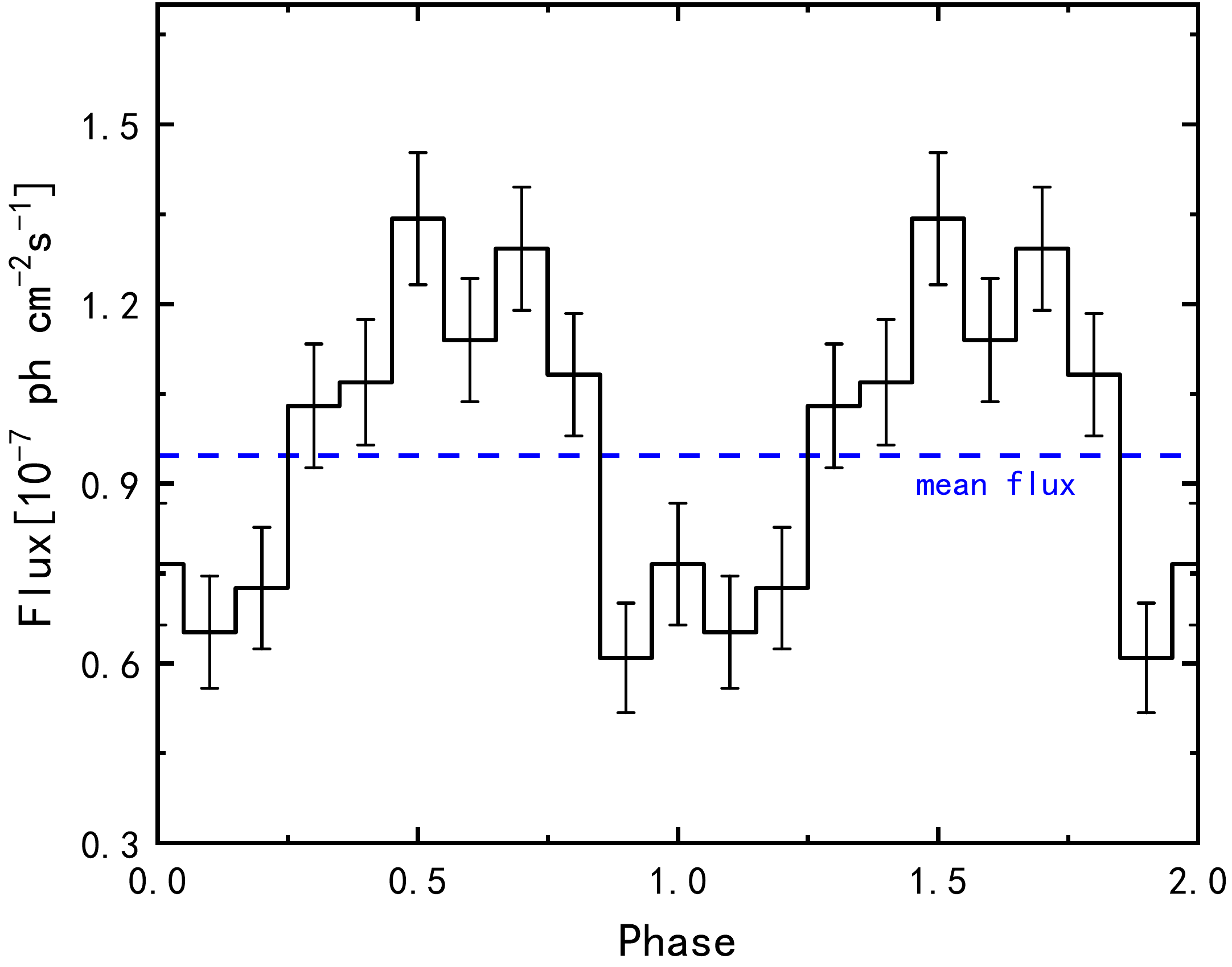}
\end{minipage}
\begin{minipage}[t]{0.49\textwidth}
\centering
\includegraphics[height=6.5cm,width=7.8cm]{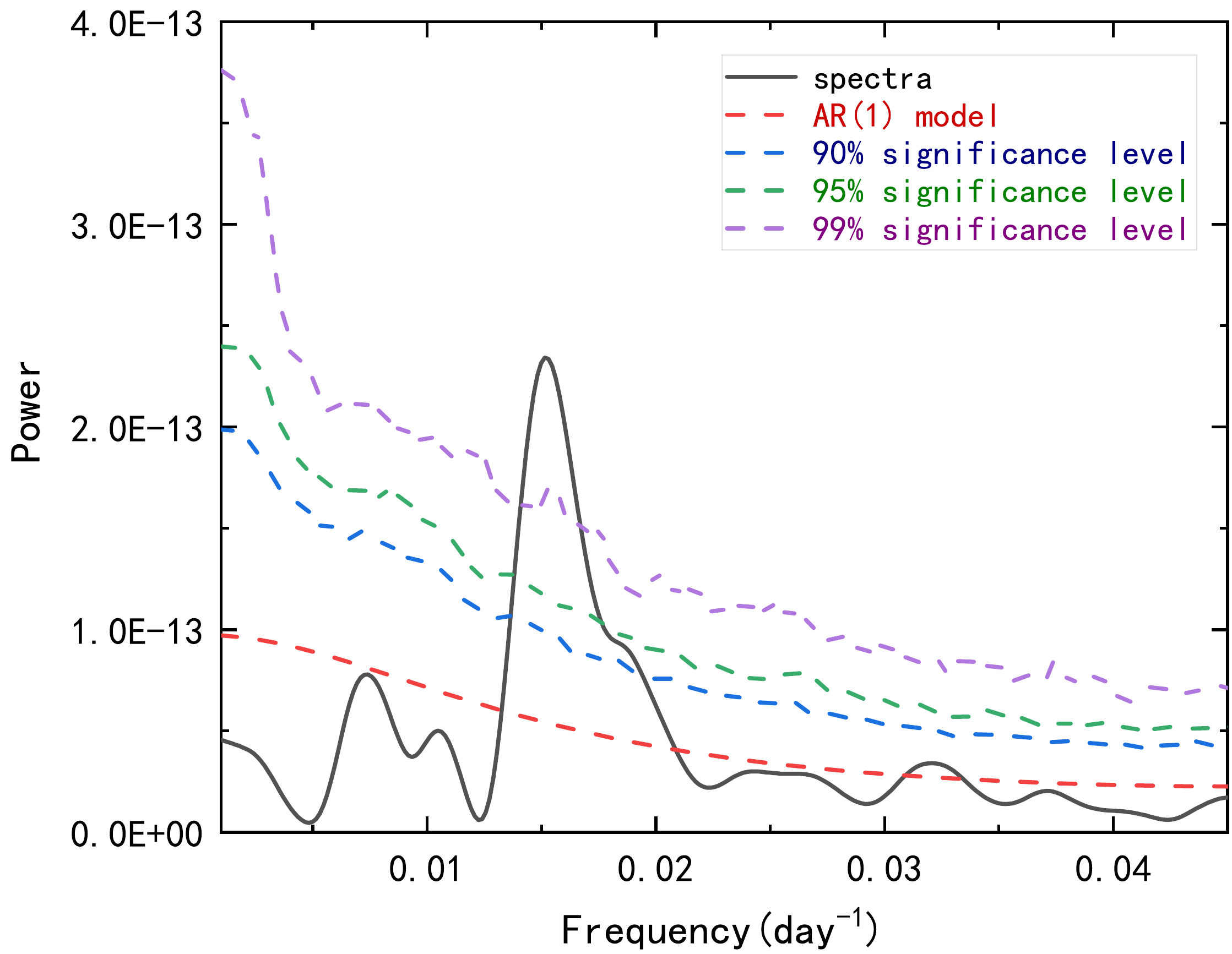}
\end{minipage}
\begin{minipage}[t]{0.99\textwidth}
\centering
\includegraphics[height=8cm,width=13cm]{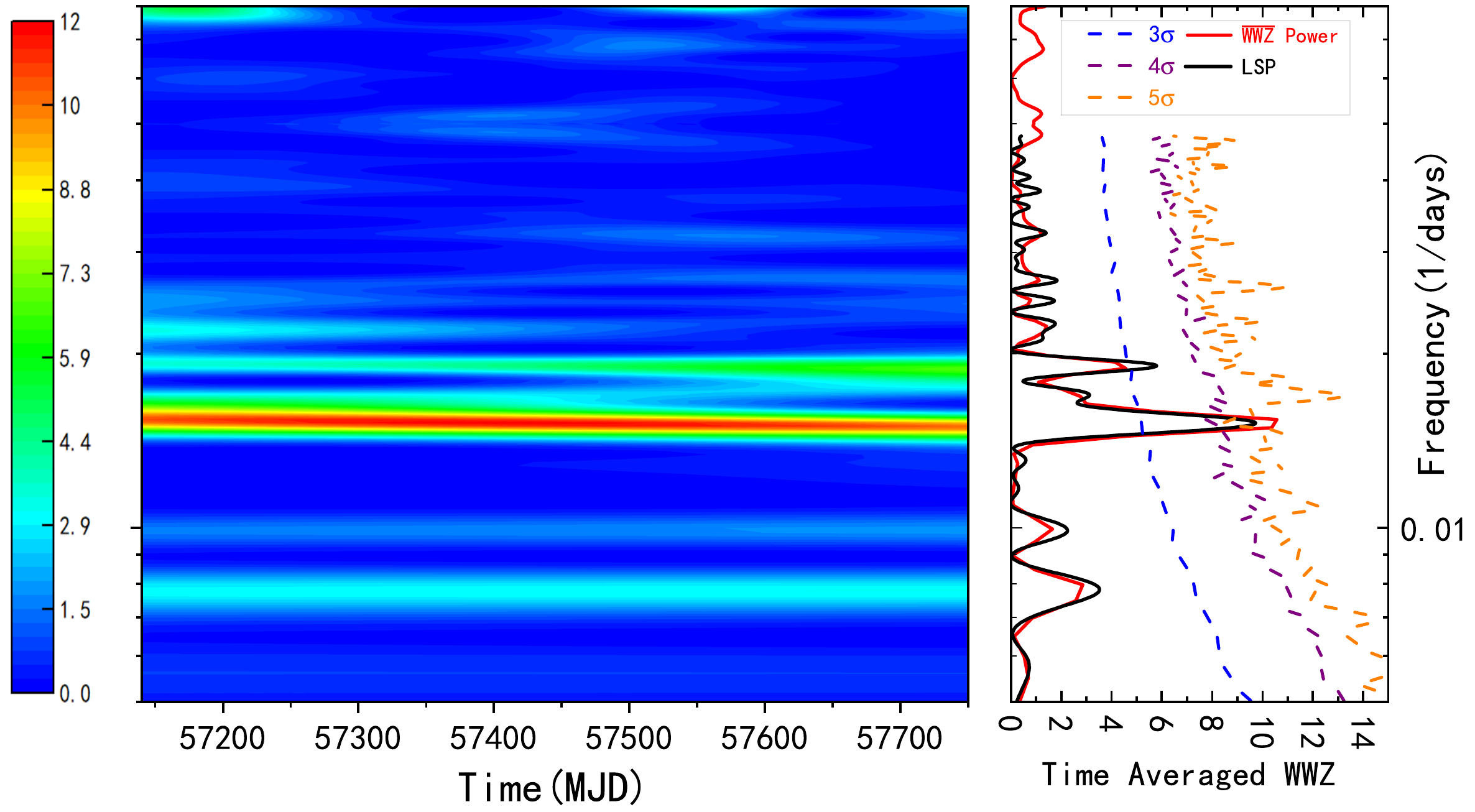}
\end{minipage}
\caption{The results of QPO analysis in segment 1 (MJD 57145--57745). Upper-left: The folded \gr\ light curve, which is constructed from binned likelihood analysis of the nine cycles of segment 1. Phase zero corresponds to MJD 57145 and 10 phase ranges are set. For clarity, we show two period cycles. The dashed blue horizontal line represents the mean flux. Upper-right: The periodicity analysis results from REDFIT. The solid black line indicates the bias-corrected power spectrum. The red dashed line represents the theoretical AR(1) spectrum. The blue, green, and purple dashed curves were 90\%, 95\%, and 99\% confidence contours, respectively. Bottom: LSP and WWZ results of the \gr\ time series data. The left sub-panel displays two-dimensional contour map of the WWZ power spectrum and the horizontal red patch indicates a strong QPO signal of $\sim$66 days. The right sub-panel shows the time-averaged WWZ (red solid line) as well as the LSP powers (black solid line). The blue, purple, and orange dashed curves were 3$\sigma$, 4$\sigma$, and 5$\sigma$ significance line, respectively. The dominant period of $\sim$66 d can be clearly seen crosses the 5$\sigma$ significance curve.  }
\label{Fig2}
\end{figure*}

\section{Periodicity search for \gr\ emission}
\label{results}

It is not rigorous enough to visually measure the QPO variability in the unevenly sampled light curve, but many methods have been proposed to detect periodic components and corresponding significance levels. Here, we applied four methods to analyse the light curves, i.e., the epoch folding, REDFIT, Lomb-Scargle periodogram (LSP), and WWZ. And a tool used to determine confidence levels as light-curve simulations. Although the \gr\ light curve obtained by us is evenly binned, we only consider the data points with TS $>$ 9, resulting in uneven sampling of data.

\begin{figure*}
\centering
\begin{minipage}[t]{0.49\textwidth}
\centering
\includegraphics[height=6.5cm,width=7.8cm]{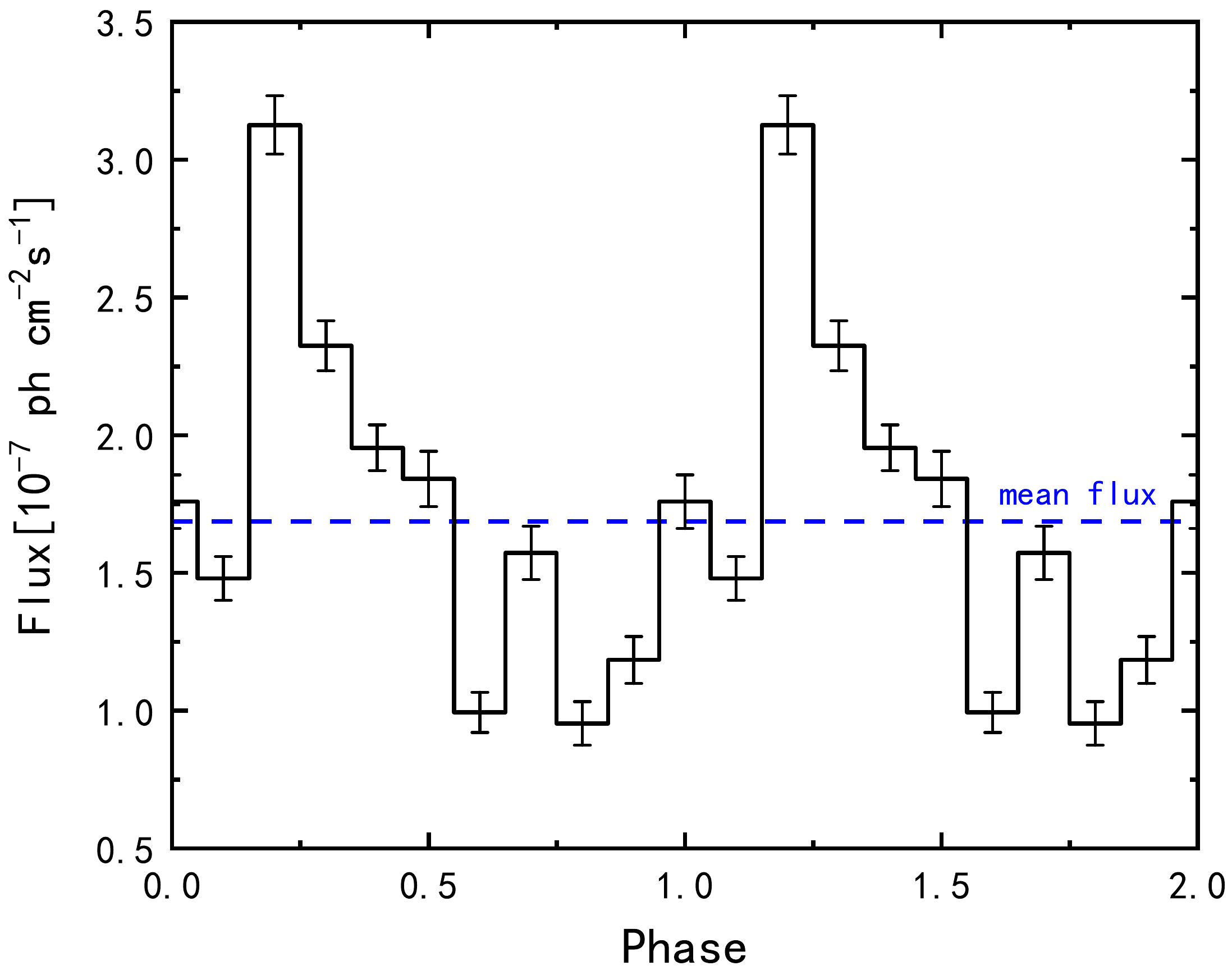}
\end{minipage}
\begin{minipage}[t]{0.49\textwidth}
\centering
\includegraphics[height=6.5cm,width=7.8cm]{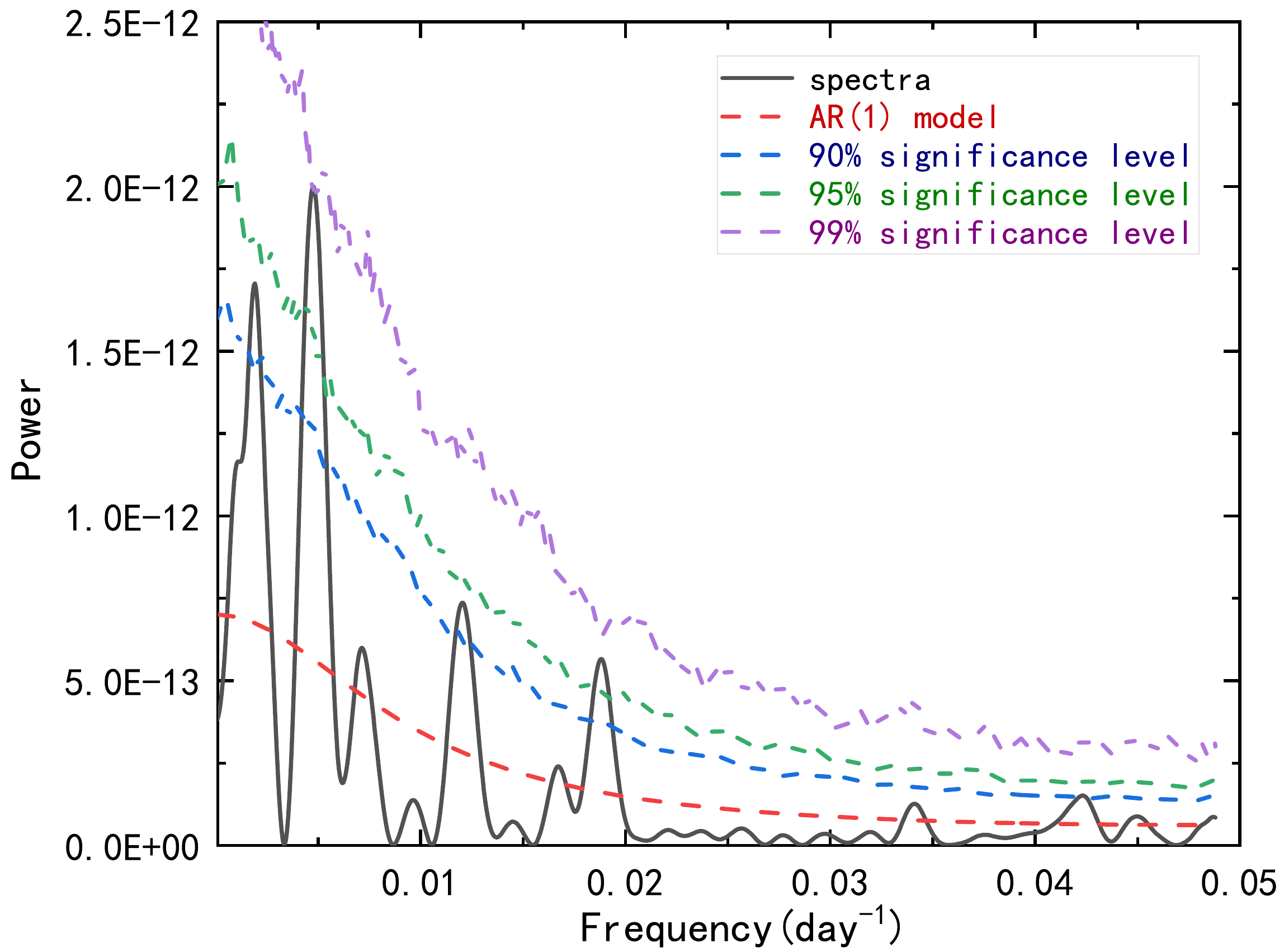}
\end{minipage}
\begin{minipage}[t]{0.99\textwidth}
\centering
\includegraphics[height=8cm,width=13cm]{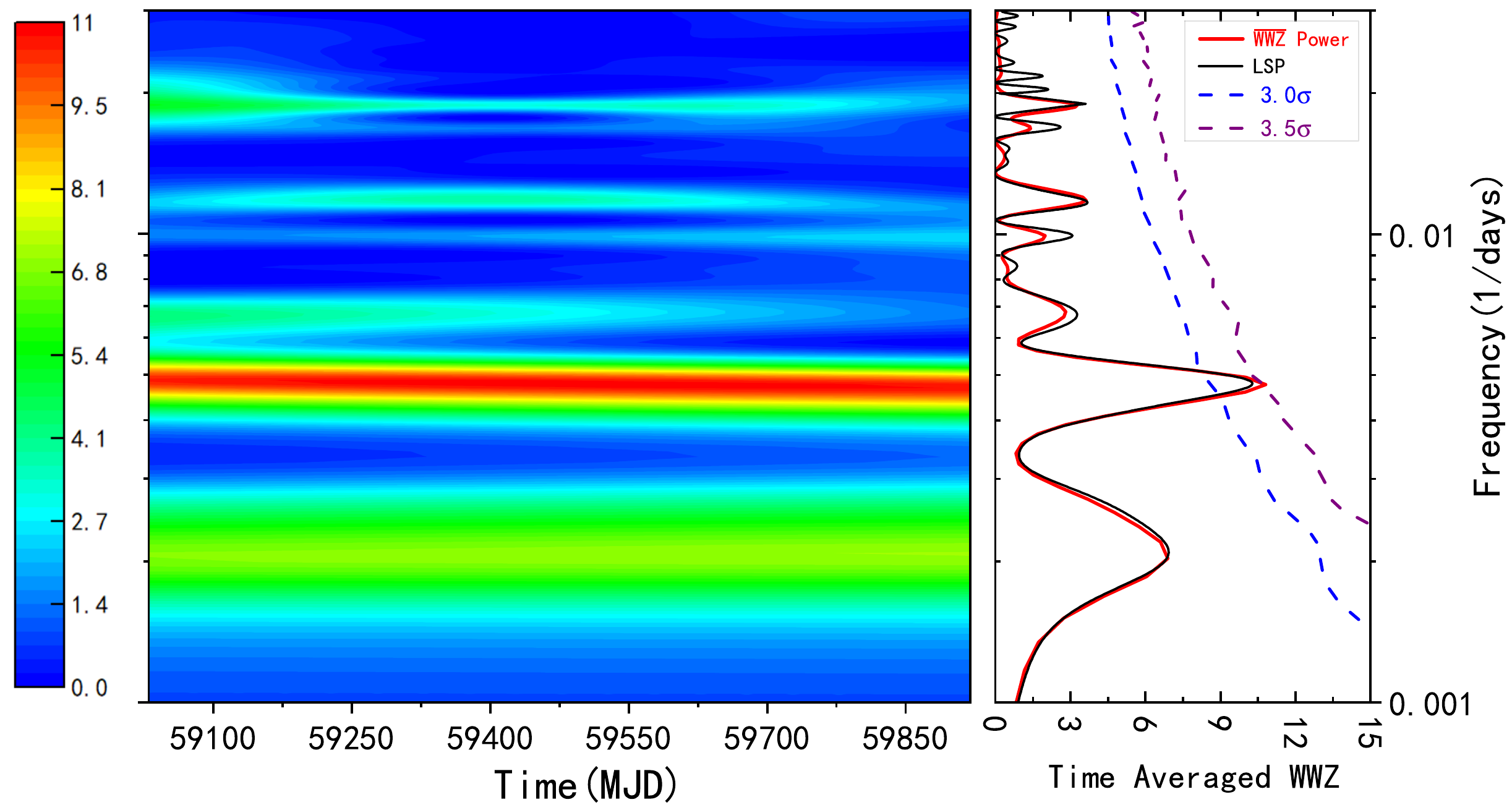}
\end{minipage}
\caption{Same as Fig~\ref{Fig2}, but for the segment 2 light curve (MJD 59035--59915). }
\label{Fig3}
\end{figure*}

The epoch folding is one of the most popular methods of light curves analysis \citep{1983ApJ...272..256L,1991MNRAS.251P..64D}. This method is insensitive to the modulating shape of periodic components and the uneven sampling of time series data, which is different from the traditional discrete Fourier periodogram \citep{Bhatta2018}. We computed $\chi^2$ values of the $\gamma$-ray light curve with a time step of 6 days for the trial periods ranging between 6 and 510 days using Equation 1 of \cite{Bhatta2018}. The results show that maximum $\chi^2$ values of 225 and 172 correspond to the trial period of 66 d and 210 d, respectively. In segment 1, we constructed a folded the light curve by the binned likelihood analysis with a $\sim66$ d period, where phase zero corresponds to MJD 57145 and 10 phase ranges are selected (the upper left panel of Figure~\ref{Fig2}). Similar to segment 1, the phase zero of segment 2 is set at MJD 59035 to complete the folding light curve with a $\sim210$ d period (the upper left panel of Figure~\ref{Fig3}). Both results show that the \gr\ flux varies with phase is obvious.

An additional method, REDFIT is also used to calculate the bias-corrected power-spectrum of the light curve and estimate the significance level of the corresponding dominant period \citep{2002CG.....28..421S}. This method can calculate the underlying red-noise spectrum by fitting the time series with a first-order autoregressive process (AR1), which is caused by some stochastic processes in the accretion disc or jet for blazars \citep{2014ApJS..213...26F,2019MNRAS.482.1270C}. The AR models believes that the present emission is connected with the past emission. The theoretical power spectrum of an AR1 model is given as,
\begin{equation}
    G_{rr}(f_j) = G_0\frac{1-\theta^2}{1-2\theta cos(\pi f_j/f_{Nyq})+\theta^2},
\end{equation}
where $G_0$ is the average spectral amplitude, $\theta$ is the average autoregression coefficient, and $f_j$ represents the discrete frequency up to the Nyquist frequency ($f_{Nyq}$). We used the REDFIT3.8e program to estimates the power spectrum and the significance level of the corresponding peak based on LSP in combination with Welch overlapped segment averaging \citep{1967ITAE...15...70W}. As can be seen from the upper right panel of Figure~\ref{Fig2}, it is evident that there is a peak around the timescale of 65 $\pm$ 12 days with significance level of $>$99\% in the power-spectrum during MJD 57145--57745 (segment 1). The upper right panel of Figure~\ref{Fig3} shows that the periodic modulation in MJD 59035--59915 (segment 2) is centered at 210 $\pm$ 55 days with significance level of $\sim$99\%. We take the half-width at half-maximum (HWHM) of the power peak fitted by the Gaussian function as the uncertainty of the periodic modulation signal.

The Lomb-Scargle periodogram (LSP) is one of the most common methods to find periodicities in time series with non-uniform sampling, and it can calculate the power spectrum intensity at different frequencies \citep{Lomb1976,1982ApJ...263..835S}. This method is the projection of the light curve on sinusoidal functions and constructs a periodogram from the goodness of the weighted $\chi^2$ fit statistic \citep{1981AJ.....86..619F}. Nevertheless, the aperiodic part of time series data will reduce the goodness of LSP sinusoid fit, which leads to the reduction of transient periodic power. The bottom panel of Fig.~\ref{Fig2} shows the power (black solid line) of the LSP for the extracted segment 1 data. One signal, at the period of 66 $\pm$ 4.8 days, reached that significance level. Meanwhile, the bottom panel of Fig.~\ref{Fig3} also shows the analysis result of segment 2. The analysis revealed a significant signal centred at 208 $\pm$ 43 day.

Further evidence for the two transient QPO is presented in the WWZ method. The WWZ method introduced by \cite{Foster1996} for the first time can identify the localized features in both time and frequency domains, especially in unequally spaced data, based on three trial functions, i.e., $\phi_1(t) = 1(t)$, $\phi_2(t) = cos[\omega(t-\tau)]$ and $\phi_3(t) = sin[\omega(t-\tau)]$. The calculation of WWZ power intensity can search for periodic modulation signal with frequency $\omega$ and time shift $\tau$ in a statistical manner, which is described as:
\begin{equation}
    WWZ = \frac{(N_{eff}-3)V_y}{2(V_x-V_y)},
\end{equation}
where $N_{eff}$ denote the effective number density of data points contributing to the signal, and $V_x$ and $V_y$ are the weighted variations of the non-uniform data $x$ and the model function $y$, respectively. For more details on the definition of these factors, see \cite{2021MNRAS.506.1540L} and references therein. For segment 1, we set the frequency range from 0.005 to 0.08 $d^{-1}$ and the step size is 0.00005 $d^{-1}$ in WWZ analysis, which enables the QPO timescale of the region of interest to be displayed as much as possible. Furthermore, in order to balance the frequency and time resolution, we set a decay constant of c = 0.001. The colour-scaled WWZ power of the 10-d binned light curve in the time-period plane are presented in the bottom panel of Fig.~\ref{Fig2}, which shows that the power for the characteristic period centred around 66 days persist the entire observational period. The corresponding time-averaged WWZ power is centred at the periods of 66 $\pm$ 4.7 day, corroborating the LSP result. In segment 2, we adopted a limited frequency range of 0.001-0.03 $d^{-1}$ in WWZ analysis, where the step size and decay constant are the same as segment 1. As shown in the bottom panel of Fig.~\ref{Fig3}, the time-averaged WWZ power of segment 2 light curve also shows a significant peak lasting throughout the activity at 208 $\pm$ 40 day, which is similar to the feature of LSP analysis.

The flux variability of blazars usually shows a frequency dependent colored-noise-like behavior, which is very likely to lead to pseudo period in the identification of periodic components of time series data, especially at lower temporal frequencies \citep{2003MNRAS.345.1271V,2016MNRAS.461.3145V,2016ApJ...832...47B,Li2017}. The significance estimation of REDFIT method is based on the $\chi^2$ distribution of periodogram points about the model, which can avoid underestimating the significance of power spectral density (PSD) peak. Here, the significance of segment 1 and 2 obtained by using the REDFIT method reveals a $\ge$ 99\% level. Another way to estimate the significance in the LSP and WWZ peaks is to simulate light curves with the same PSD and flux distribution as the original light curve using a Monte Carlo method provided in \cite{2013MNRAS.433..907E}. The underlying red-noise PSDs of blazar light curves are often reasonably approximated to a power-law form $P(f) \propto f^{-\alpha}$, where $P(f)$ is the power at temporal frequency $f$ and $\alpha$ is spectral slope \citep{2005A&A...431..391V}. Then, we generated $10^6$ artificial light curves to estimate the significance level of the LSP and WWZ periodic components. In segment 1, the significance level for the QPO signal was found to be $> 5\sigma$ (the bottom panel of Fig.~\ref{Fig2}). In segment 2, the light curve simulation shows that the periodic modulation of 210 d seems to have a significance level close to $3.5\sigma$ (the bottom panel of Fig.~\ref{Fig3}). In the recent QPO search, a large number of blazars claimed to have periodic signals are usually greater than $3\sigma$ significance \citep{2020ApJ...896..134P,2020ApJ...891..163Z,2021ApJ...919...58Z}. Thus, the QPO signal with $\sim3.5\sigma$ significance of segment 2 is sufficiently important to be reported. These two transient QPO signals may again or continue to appear in the future, so it will be interesting to keep monitoring at the $\gamma$-ray frequency.

\section{Conclusions and Discussion}
\label{sumdis}
We collected 0.1-300 GeV energy band data of the blazar S4 0954+658 from the Fermi-LAT archive and conducted a temporal analysis in two interesting periods: segment 1 (MJD 57145--57745) and segment 2 (MJD 59035--59915). Four analytical methods (e.g., the epoch folding, REDFIT, LSP, and WWZ) and a tool (light curve simulations) are called to detect the transient QPO in the 10-d binned light curve, revealing a good consistency between different methods. For segment 1, our results showed that there was a 66 day QPO above $5\sigma$ significance level during MJD 57145--57745, which lasted for nine cycles. Interestingly, the 66 day periodic modulation, similar to the PKS 2247-131 case, also occurred after a outburst event (2014 December) with multi-wavelength observations \citep[See Fig.~\ref{Fig1};][]{Zhou2018,2019MNRAS.484.5633G}. For segment 2, we found a possible QPO of about 210 day with $\sim3.5\sigma$ significance in the over 880 day \gr\ light curve. This signal is clearly visible for about four cycles and seems to continue to appear after MJD 59915 (2022 December). It is of interest to keep monitoring the source, checking whether or not the QPO signal of $\sim$210 day would appear again. Unfortunately, we can not verify the authenticity of the two transient QPOs in the multi-wavelength light curve due to lack of good coverage of multi-wavelength observations and data point resolution during the concerned period. We expect that different telescopes will pay attention to the QPO signal of this source in the future.

A variety of scenarios have been proposed to explain the QPO phenomenon in blazar emission. One of the most interesting features of the accretion flow are the stable twin high-frequency QPO often appear with frequency ratio 3:2 in the X-ray flux, e.g., Sgr A$^*$ and GRO J1655-40 \citep{2001A&A...374L..19A,2005A&A...440....1T}. Two stable peaks QPOs scale indicates that they can originate from some resonant process taking place in the accretion disk's oscillations \citep{2003A&A...404L..21A,2009A&A...499..535H}. In the framework of the resonance model, the frequencies reflect epicyclic motion of perturbed flow lines in the accretion disc, or combinations between these and a fixed, perturbation frequency \citep{2005MNRAS.357L..31R}. The scaled similarity between stellar mass systems and AGNs indicates that resonances are important for AGNs as well. Although no pairs of QPOs at that 3:2 ratio have been detected for S4 0954+658, as 66 d and 210 d correspond to frequencies of $1.75 \times 10^{-7}$ Hz and $0.55 \times 10^{-7}$ Hz, respectively. Separate but related is the relativistic precession model, which associates three different QPOs to a combination of the fundamental frequencies of particle motion \citep{2014MNRAS.439L..65M}. While the higher frequency QPOs correspond to the Keplerian frequency of the innermost disk regions, the lower frequency QPOs correspond to the relativistic periastron precession of eccentric orbits and the Type-C QPOs in the nodal precession (or Lense-Thirring precession) of tilted orbits in the same regions \citep{1998ApJ...492L..59S,1999ApJ...524L..63S}. For the Lense-Thirring precession, the period can be expressed using $\tau_{LT} = 0.18a_s^{-1}(M/10^9M_{\odot})(r/r_g)^3 $days, where $a_s$, $M$, $r_g$ and $r$ is dimensionless spin parameter, mass of the black holes (BH), the gravitational radii and the radial distance of the emission region from the BH, respectively. In such scenario, taking the spin parameter $a_s = 0.9$ and the BH mass $M = 2.3 \times 10^{8} M_{\odot}$ \citep{2021MNRAS.504.5258B}, the timescale of the two QPOs places the emission region ranges from 10 to 15 $r_g$. Due to the warped accretion discs, the QPO phenomenon could be the result of the jet precession, therefore resulting in a periodic timescale of thousands years \citep{2018MNRAS.474L..81L,Bhatta2018,2023MNRAS.519.4893L}. Such a large timescale does not seem to apply to this case.

A binary SMBH system model was proposed to explain the $\sim$2 yr periodic fluctuation in the multiwavelength light curve of PG 1553+113 and later applied to interpret the similar fluctuation behavior of other blazers \citep{2015ApJ...813L..41A,2018A&A...615A.118S,2020MNRAS.492.5524O,2022ApJ...929..130W}. The orbital motion of this model may cause a long-term periodic temporal signals, which is reflected in the periodic accretion perturbations, or jet-precessional and nutational motions \cite{2018MNRAS.474L..81L}. The observed period $P_{\mathrm{obs}}$ is corrected to the intrinsic orbital period at the local galaxy via the relation $P_{\mathrm{int}} = P_{\mathrm{obs}}/(1+z)$, where $z = 0.3694$ is the cosmological redshift. Using the period values of 66 d and 210 d, we get the intrinsic orbital period values of 48 d and 153 d respectively. Assuming that the mass ratio between the two SMBHs is 0.1 and taking the mass of the central black hole to be $2.3 \times 10^{8} M_{\odot}$ as that of the primary black hole. We substitute two transient QPO values into the formula given by \cite{2010RAA....10.1100F}, and the results show a very tight orbit (0.001 pc and 0.002 pc) and a quick merging timescale (95 yr and 2048 yr) in the gravitational waves driven regime \citep{Bhatta2018}. Nevertheless, the two transient QPOs we detected were too short compared to the periodic timescale expected by this model. And a binary SMBH system should produce a more stable/persistent periodic behaviour which is not observed.

Another potential explanation for the two transient QPOs is a geometrical model with plasma blobs moving helically down the jet, which has been recently applied in many cases \citep{Zhou2018,2021MNRAS.506.1540L,2022MNRAS.510.3641R}. In this model, as the plasma blob (contains higher particle and magnetic energy densities) that injected into jet enhances the emission, every plasma blob will change its orientation with respect to the line of sight and this will produce a quasi-periodic flux modulation due to the Doppler beaming effect. The plasma blobs moving helically within the jet may be a natural process in magnetically dominated jets \citep{2021ApJ...906..105C}. In the helical motion of the blob, $\theta_{obs}(t)$ of a given emitting region depends on the pitch angle of the helix $\phi$ and on the angle $\psi$ of the axis of the jet with respect to our line of sight according to,
\begin{equation}
    \cos \; \theta_{obs}(t) = \cos\;\phi \; \cos\;\psi + \sin\;\phi \; \sin\;\psi \; \cos \; \omega (t).
\end{equation}
where $\omega(t) = 2\pi t/P_{obs}$ is the variable azimuthal and $P_{obs}$ is the observed period. From $\theta_{obs}(t)$, and adopting the bulk Lorentz factor $\Gamma$ = 11.4 given by \cite{2017ApJ...846...98J}, we calculate the Doppler factor $\delta(t)$ with equation, $\delta(t) = 1/\Gamma(1-\beta cos \; \theta_{obs}(t))$, where $\beta = \nu_{jet}/c$. Then, the periodicity in the blob rest frame can be calculated (see \cite{2022MNRAS.510.3641R} for details). For the case of S4 0954+658, if we assume the parameters used in \cite{2017ApJ...846...98J} for the parsec-scale radio jet, the pitch angle $\phi = 1.75^\circ$ (assumed to be half of the opening angle), the
viewing angle $\psi = 1.5^\circ$, and $P_{obs} = 66$ d, the blob traverses about a distance $D = 9c\beta \, P_{\mathrm{rest}} \; \cos \phi \; \sin \psi \approx 1.64$ pc down the jet during nine period \citep{Zhou2018}. In addition, for $P_{obs} = 210$ d, the blob travels $\sim$2.32 pc during four period. As the blob is injected into the jet (or dissipates), the periodic modulation tends to become (or less) noticeable. This model has a defect that it can only explain a QPO with almost constant amplitude. However, the amplitude of the QPO is almost constant either in segment 1 or in segment 2, although it is different in the two segments (see Fig.~\ref{Fig1}). Hence, it is reasonable that the different plasma blobs of this model are used to explain the transient properties of QPO with different timescales. We expect the 210 day QPO behavior will continue to appear in Fermi-LAT observation. Furthermore, we also hope that the multi-wavelength campaign (i.e. TESS and WEBT) will pay attention to whether two transient QPOs variability also appears and identify the underlying physical mechanism among different hypotheses.
\section*{Acknowledgements}
{We thank anonymous referee for very helpful suggestions.
This research or product makes use of public data provided by Fermi-LAT.
JF is partially supported by National Natural Science
Foundation of China (NSFC) under grant U2031107, the Joint Foundation of Department of Science and Technology of Yunnan Province and Yunnan University  (202201BF070001-020), the grant from Yunnan Province (YNWR-QNBJ-2018-049)
and the National Key R\&D Program of China under grant (No.2018YFA0404204). Y.L.G. is supported by Yunnan University Graduate Scientific Research Innovation Fund under grant KC-2222975. T.F.Y. is supported by NSFC under grant 11863007.
}

\end{document}